\documentstyle[times,pramana,epsf,floats]{ias}
\begin{document}
\title{Direct Photon Production in Heavy Ion Reactions at SPS and RHIC}

\author{T. Peitzmann}
\address{Institut f\"ur Kernphysik, University of M\"unster, 48149 
M\"unster, Germany}
\abstract{A review on experimental results for direct photon 
production in heavy ion reactions is given. A brief survey of early 
direct photon limits from SPS experiments is presented. The first 
measurement of direct photons in heavy ion reactions from the WA98 
collaboration is discussed and compared to theoretical calculations. 
An outlook on the perspective of photon measurements at RHIC is given.
}

\maketitle
\section{Introduction}
The major motivation to study relativistic heavy-ion collisions is
the search for the quark-gluon plasma (QGP), a potential new state of 
matter where
colored quarks and gluons are no longer confined into hadrons and
chiral symmetry is restored. 
To study such a complicated system one wishes for a probe that is not
equally complicated in itself. The production of hadrons is of course
governed by the strong interaction and therefore adds to the
complication.  One possible way out might be the study of hard
processes where QCD, the theory of strong interaction, enters the
perturbative regime and is calculable. The other avenue involves a
particle that suffers only electromagnetic interaction: Photons ---
both real and virtual --- should be an ideal 
probe.\footnote{For previous reviews on this topic see Refs. 
\protect\cite{Alam96,Alam01,Pei02}.}
While photon production may be less
difficult to treat than some other processes in hadronic physics, an
adequate treatment in heavy-ion collisions turns out to be far from
trivial.
Experimentally, high energy direct photon measurement has always been 
considered
a challenge. This is true already in particle physics and even more in
the environment of heavy-ion collisions. Nevertheless a lot of
progress has been made and a large amount of experimental data is
available, though mostly from particle physics. 
Only one measurement of direct photons exists for heavy ion 
collisions and was recently published by the WA98 collaboration 
\cite{Aggarwal:2000}.
Direct photon measurements
in heavy-ion collisions are expected to come into real fruition with
the advent of colliders like RHIC and LHC.
In the present report I attempt to provide a review of
the experimental aspects of the study of direct
photon production in heavy-ion collisions. I will first present 
results from the CERN SPS fixed target program and comparisons to 
theoretical calculations. The second part will discuss the 
experimental potential of direct photon measurements at RHIC.

\section{Experimental results at SPS}

In heavy-ion collisions the extraction of direct photons is extremely 
difficult due to the  
high particle multiplicity. The highest available energy in heavy-ion 
collisions so far at the CERN SPS has been approximately at the lowest energy 
where direct photons could be measured in $pp$.

Using the relatively light ion beams of $^{16}O$ and $^{32}S$ at a 
beam energy of 200~$A$~GeV, corresponding to a nucleon-nucleon center of 
mass energy of $\sqrt{s_{NN}} = 19.4 \, \mathrm{GeV}$, the experiments 
WA80 \cite{Albrecht:1996,Albrecht:1991}, HELIOS (NA34) \cite{zpc:ake:90} 
and CERES (NA45) \cite{zpc:bau:96} have attempted to measure direct 
photons. All these measurements have been able to deliver upper 
limits of direct photon production. 

HELIOS has studied $p$-, $^{16}O$- and $^{32}S$-induced reactions 
\cite{zpc:ake:90} with a conversion method. 
The authors estimate the 
ratio of the integrated yields of inclusive photons and neutral pions:
\begin{equation}
    r_{\gamma} = \frac{N_{\gamma}}{N_{\pi^{0}}}
    \label{eq:helios}
\end{equation}
for $p_{T} > 100 \, \mathrm{MeV}/c$. They calculate the neutral 
pion yield from the number of negative tracks in their magnetic 
spectrometer. Their results (with $4-11 \% $ 
statistical and $9 \% $ systematic uncertainty) and their estimate of decay 
photons (with $9 \% $ systematic uncertainty) agree within these errors. 
An analysis of the $^{32}S$-induced data with a higher cutoff of 
$p_{T} = 600 \, \mathrm{MeV}/c$ yields a comparable result. However, 
the results are of limited value in the context of both prompt and thermal 
direct photons, as they are dominated by the lowest $p_{T}$, where the 
expected direct photon emission would be negligible.

\begin{figure}[tb]
    \centering
    \epsfbox{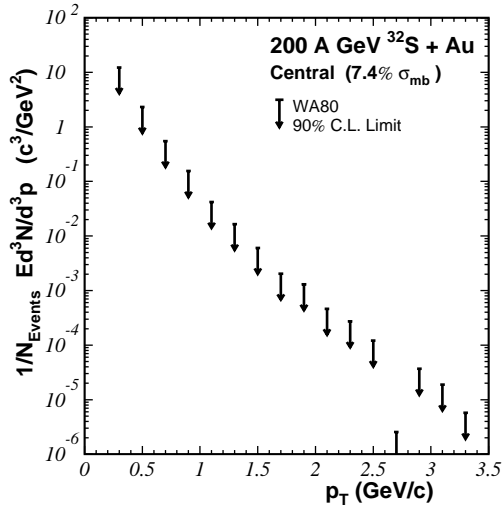}
    \caption{Upper limits ($90 \% $ CL) of the direct photon 
    multiplicity as a function of  $p_{T}$ in central 
    reactions of $^{32}S$~+~$Au$ for 
    $\sqrt{s} = 19.4 \, \mathrm{GeV}$.}
    \label{fig:wa80}
\end{figure}

\begin{figure}[tb]
    \centering
    \epsfbox{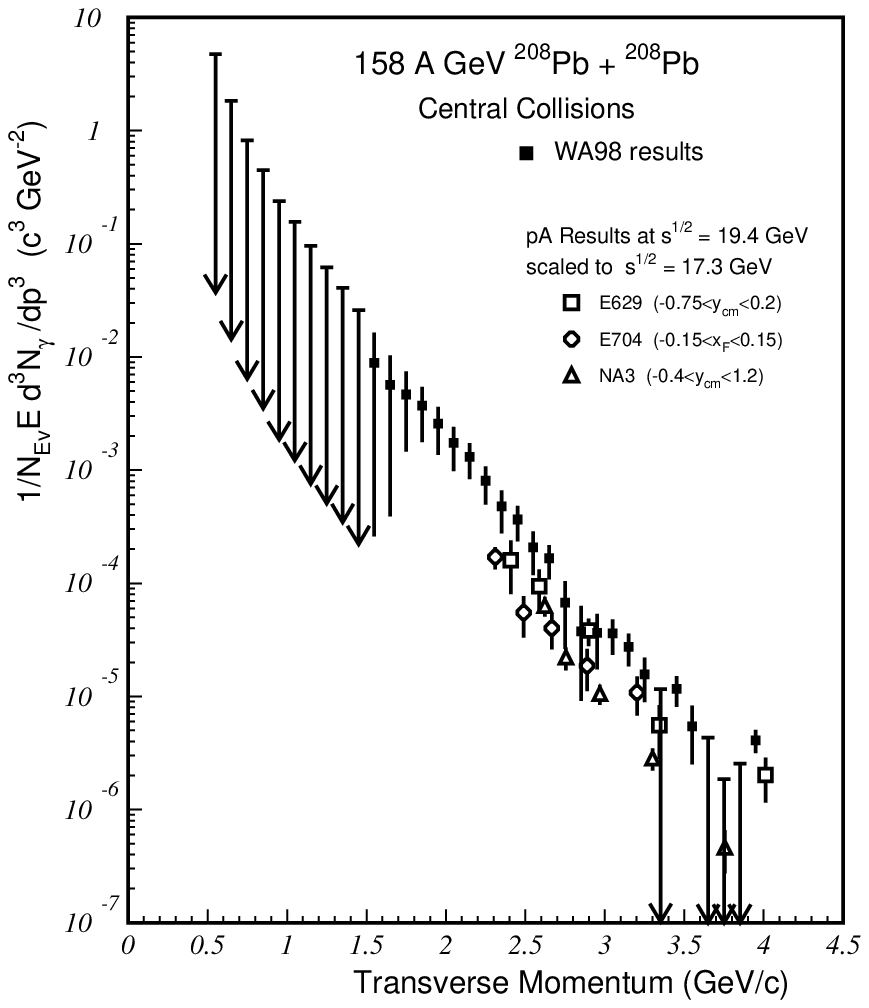}
    \caption{Invariant direct photon 
    multiplicity as a function of  $p_{T}$ in central 
    reactions of $Pb$~+~$Pb$ for 
    $\sqrt{s} = 17.3 \, \mathrm{GeV}$. The error bars correspond to 
    combined statistical and systematic errors, the data points with downward 
    arrows indicate $90 \% $ CL upper limits. For comparison scaled direct 
    photon results from $p$-induced reaction are included (see text).}
    \label{fig:wa98dat}
\end{figure}

A similar measurement has been performed by the CERES experiment, 
which has studied $^{32}S$~+~$Au$ reactions \cite{zpc:bau:96}. 
Photons are measured when 
they convert in the target, the e$^{+}$-e$^{-}$-pairs are 
reconstructed by tracking in the two RICH detectors. 
They obtain 
inclusive photon spectra in central $^{32}S$~+~$Au$ reactions in
$0.2 \, \mathrm{GeV}/c \le p_{T} \le 2.0 \, \mathrm{GeV}/c$. The 
results agree within errors with their hadron decay generator, which 
is tuned to reproduce charged and neutral pion spectra from different 
heavy-ion experiments. They estimate a similar ratio of integrated 
yields:
\begin{equation}
    r_{\gamma}^{\prime} = \left( \frac{dN_{ch}}{d\eta} \right)^{-1}
    \int_{0.4 \, \mathrm{GeV}/c}^{2.0 \, \mathrm{GeV}/c}
    \frac{dN_{\gamma}}{dp_{T}}dp_{T},
    \label{eq:ceres}
\end{equation}
which they use --- again by comparing to the generator 
--- to establish an upper limit ($90 \% $ CL) of $14 \% $ for 
the contribution of direct photons to the integrated inclusive photon 
yield. One of the uncertainties which is difficult to control in 
this analysis relates to the fact that they use simulated hadron 
yields in their generator which are tuned to other measurements with 
different trigger biases and systematic errors, and that especially the 
neutral pions have not been measured within the same data set.

In addition, the CERES experiment has utilized another method to extract 
information on a possible direct photon contribution. As in 
na{\"{\i}}ve pictures of particle production in these reactions the 
direct photon multiplicity is proportional to the square of the 
initial multiplicity while the hadron multiplicity should be 
proportional to the initial multiplicity, they have studied the 
multiplicity dependence of the inclusive photon production. Their 
upper limit on a possible quadratic contribution is slightly lower 
than the above limit on direct photons from $r_{\gamma}^{\prime}$, 
its relation to the direct photon contribution 
is however dependent on the model of particle production.
Similar to the HELIOS measurements both these results are dominated 
by the low $p_{T}$ part of the spectra, so the result is consistent 
with the expectation of a very low direct photon yield at low $p_{T}$.

The WA80 experiment has performed measurements with 
$^{16}O$ \cite{Albrecht:1991} and 
$^{32}S$ \cite{Albrecht:1996} beams using a lead glass calorimeter 
for photon detection. The systematic errors are check\-ed by 
performing the analysis with a number of different choices of 
experimental cuts.
Inclusive photons and $\pi^{0}$ and $\eta$ 
mesons have been measured in the same data samples, which helps to 
control the systematic errors. WA80 reports no significant direct 
photon excess over decay sources in peripheral and central collisions 
of $^{16}O$~+~$Au$ and $^{32}S$~+~$Au$. The average excess in central 
$^{32}S$~+~$Au$ collisions in the range 
$0.5 \, \mathrm{GeV}/c \le p_{T} \le 2.5 \, \mathrm{GeV}/c$ is given 
as $5.0 \% \pm 0.8 \% $ (statistical) $\pm 5.8 \% $ (systematic). A 
$p_{T}$ dependent upper limit ($90 \% $ CL) of direct photon 
production as shown in Fig. \ref{fig:wa80} has been obtained, which 
gives more information than the integrated limits, as it can constrain 
predictions at higher $p_{T}$, where a considerable direct photon 
multiplicity may be expected.

The upper limits for direct photons from WA80 have been used by a 
number of different authors to compare their model predictions 
\cite{Srivastava94,Shuryak94,Neumann95,Arbex95,Dumitru95,Sollfrank97,Sarkar99,Cleymans97,Cleymans98,Srivastava00a}.
They can be explained
with and without phase transition and, therefore, do not allow a
conclusion about the existence of a QGP phase. However, they have
triggered investigations of some of the simplifications used in
earlier calculations, as e.g. unrealistic equations of state 
for the hadron gas.

For $Pb$~+~$Pb$ collisions at 158~$A$~GeV ($\sqrt{s_{NN}} = 17.3 \, 
\mathrm{GeV}$) the WA98 experiment has performed photon measurements
\cite{Aggarwal:2000}
using similar detectors and analysis techniques as WA80. In 
peripheral collisions no significant direct photon excess was found. 
In central collisions the observed photons cannot entirely be 
explained by decay photons, implying the first observation of direct 
photons in high energy heavy-ion collisions. The extracted direct 
photon spectrum is shown in Fig. \ref{fig:wa98dat}. The only other 
direct photon measurements at a similar energy are from $p$-induced 
reactions. Data from 
$pp$ reactions by E704 \cite{plb:ada:95} 
and from $p$+$C$ reactions by E629 \cite{prl:mcl:83} 
and NA3 \cite{zpc:bad:86} at $\sqrt{s} = 19.4 \, \mathrm{GeV}$ have 
been converted to the lower energy $\sqrt{s} = 17.3 \, \mathrm{GeV}$ 
assuming a scaling according to 
\cite{rmp:owens:87}:
\begin{equation}
    E d^3\sigma_{\gamma}/dp^3 = f(x_T,\theta)/s^2, 
    \label{eq:scaling}
\end{equation}
where $x_T=2p_T/\sqrt{s}$ and $\theta$ is the emission angle of the 
photon and have 
been multiplied with the average number of binary nucleon-nucleon 
collisions in the central $Pb$~+~$Pb$ reactions (660). These scaled 
$p$-induced results are included in Fig. \ref{fig:wa98dat} for comparison. 
They are considerably below the heavy-ion results which indicates that 
a simple scaling of prompt photons as observed in $pp$ is not sufficient 
to explain the direct photons in central $Pb$~+~$Pb$ reactions. 

\begin{figure}[tbh]
    \centering
    \epsfbox{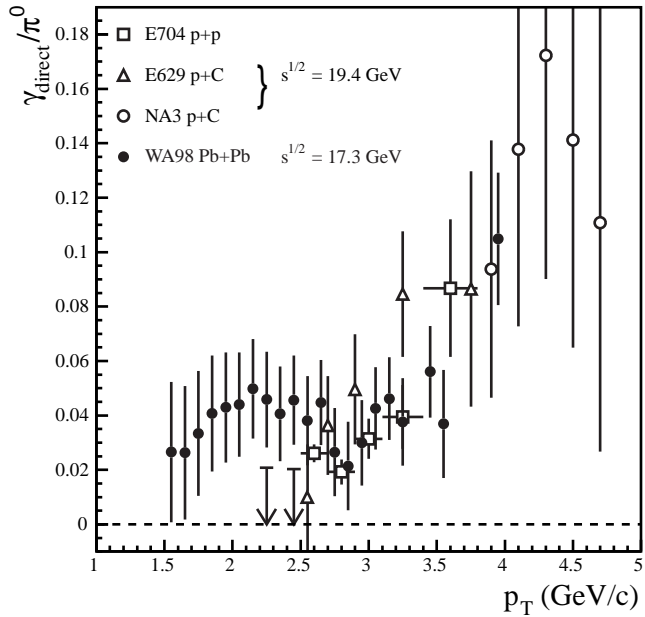}
    \caption{$\gamma_{direct} / \pi^{0}$ ratio as a function of $p_{T}$ 
    for $pp$ and $pC$ reactions
    $\sqrt{s} = 19.4 \, \mathrm{GeV}$ as in 
    Fig. \protect\ref{fig:wa98dat} (open symbols) 
    and for central $Pb$~+~$Pb$ reactions at 
    $\sqrt{s} = 17.3 \, \mathrm{GeV}$ (filled symbols).}
    \label{fig:wa98gampi}
\end{figure}

It is also instructive to compare 
the $\gamma / \pi^{0}$ ratio extracted from heavy-ion 
data to those from $pp$ and $pC$ in Fig. \ref{fig:wa98gampi}. The value 
in heavy-ion data is $\approx 3-5 \% $ in most of the $p_{T}$ range, 
which is similar to the lowest values extracted in the proton data. 
This may be taken as a hint that such levels of direct photons 
approach the feasibility limit of such measurements. Still lower 
levels will be very hard or impossible to detect. Furthermore, while 
in this ratio the heavy ion data and the proton data agree for high 
transverse momenta, there is an indication of an additional component 
at intermediate $p_{T}$ in the heavy ion data.

\begin{figure}[hbt]
\centering
\epsfbox{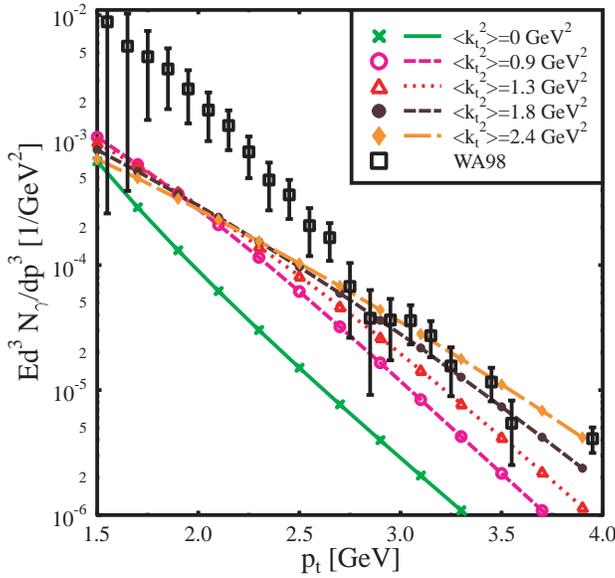}
\caption{The pQCD photon spectrum calculated for different $p_{T}$ 
broadening in comparison to the WA98 data
\protect\cite{Dumitru01}.}
\protect\label{fig4.7a}
\end{figure}

\begin{figure}[tbh]
\centering
\epsfbox{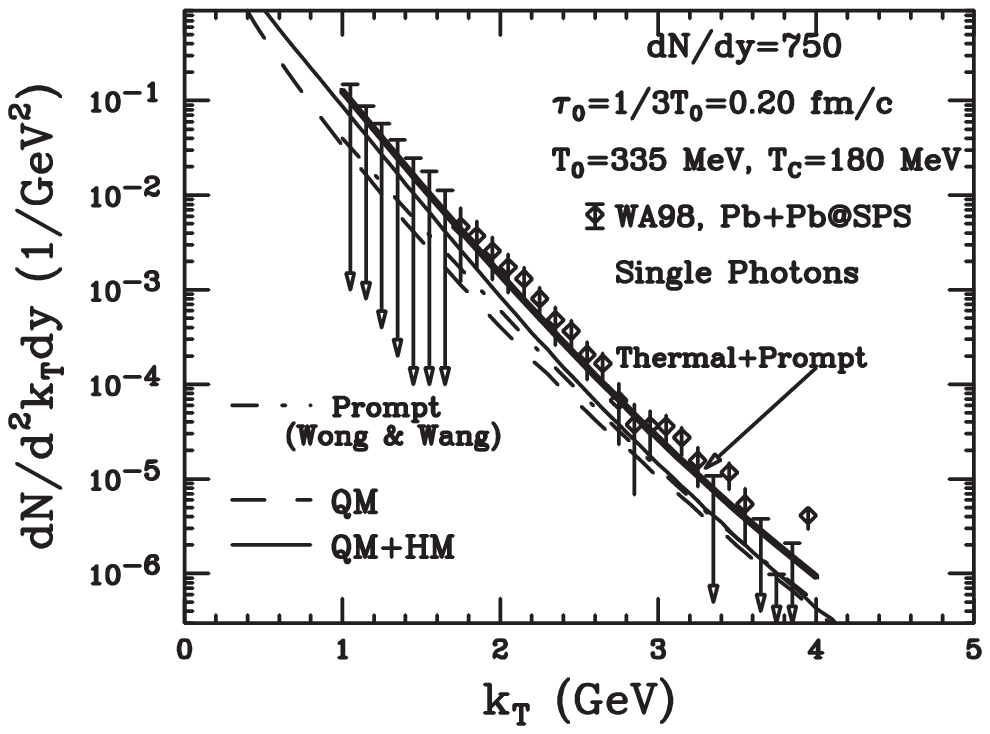}
\caption{Comparison of the WA98 data with a hydrodynamical calculation
by Srivastava and Sinha \protect\cite{Srivastava00}. The pQCD 
calculation by Wong and Wang \protect\cite{Wong98} is also shown.
\protect\label{fig4.5}}
%
\centering
\epsfbox{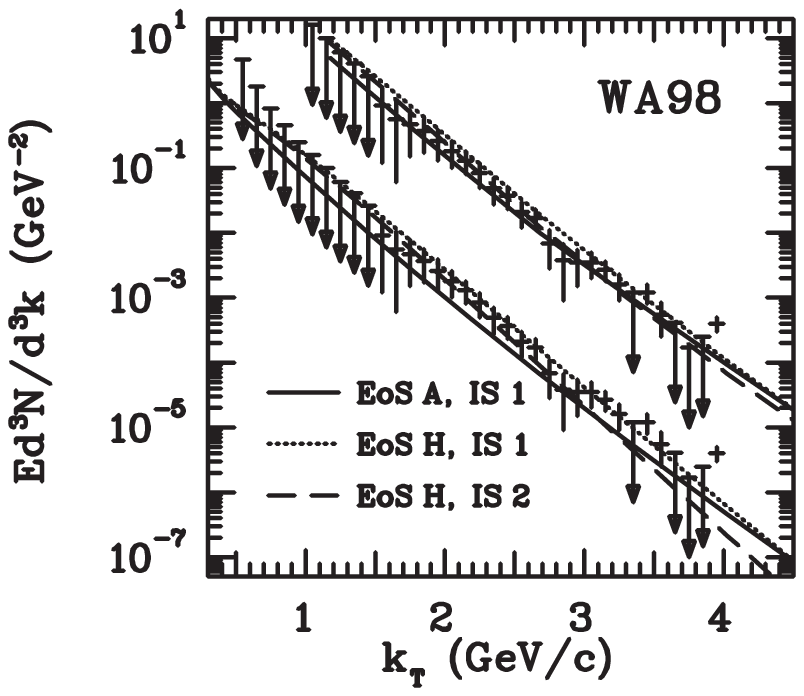}
\caption{The photon spectrum calculated for different EOS and initial 
conditions 
with prompt photons (upper set, scaled by a factor 100)
and without (lower set) in comparison to the WA98 data
\protect\cite{Huovinen00}. EoS A, IS 1 contains a phase transition at $T_c=165$
MeV, an average initial temperature $T_0=255$ MeV, and a local maximum 
temperature $T_{\rm max}=325$ MeV. EoS H describes only a HHG with
$T_0=234$ MeV, $T_{\rm max}=275$ MeV (IS 1) and $T_0=213$ MeV, 
$T_{\rm max}=245$ MeV (IS 2).
\protect\label{fig4.8}}
\end{figure}

Before attempting to address the thermal production of photons it is 
mandatory to understand the contribution from hard processes which 
are expected to dominate at high $p_{T}$.
Wong and Wang have calculated this contribution
\cite{Wong98} from a next-to-leading order perturbative QCD
calculation, where an intrinsic parton momentum of $\langle k_T^2 \rangle
=0.9$ GeV$^2$ has been used. This $\langle k_T^2 \rangle$ is 
necessary to describe p-induced reactions at a similar energy. The 
heavy ion data can, however, not be described by this calculation 
(see Fig. \ref{fig4.5}).
Dumitru et al. \cite{Dumitru01} have followed up on this question.
They showed that the WA98 photon spectrum above $p_T=2.5$ GeV
can be explained by prompt photons if an additional nuclear broadening of
$\Delta k_T^2=\langle k_T^2\rangle_{AA} -\langle k_T^2\rangle_{pp}\simeq
0.5$ - 1 GeV$^2$ is introduced. For low $p_T<2.5$ GeV, however, 
prompt photons fail to reproduce the WA98 data regardless of the amount of
nuclear broadening employed (see Fig. \ref{fig4.7a}).

A number of groups
\cite{Srivastava00,Alam01a,Peressounko00a,Gallmeister00,Huovinen00,Chaudhuri00,Steffen01}
have compared their hydrodynamical calculations with the data of 
WA98. I will only mention a few examples - for a more detailed 
discussion see \cite{Alam-icpaqgp}.

Srivastava and Sinha  argued, using the 2-loop hard thermal loop rate
for the QGP contribution and a realistic equation of state for the 
hadron gas,
that the QGP is needed to explain the data.
Their conclusion
is based on the use of a very high initial temperature ($T_0=335$ MeV)
and very small initial time ($\tau_0=0.2$ fm/c), which could explain
the observed flat photon spectrum for transverse photon momenta $p_T>2$ GeV
(see Fig. \ref{fig4.5}).
Srivastava and Sinha
have also included prompt photons from the work by Wong and Wang
\cite{Wong98}. Srivastava and Sinha found that the thermal
photons contribute half of the total photon spectrum and that in particular
at large $p_T$ most of the thermal photons are due to the QGP contribution.

Gallmeister et al.
\cite{Gallmeister00} argued that 
the low momentum part of the WA98 spectrum ($p_T<2$ GeV) is consistent 
with a thermal source, either QGP or hadron gas, 
which also describes dilepton data.
The hard part ($p_T>2$ GeV), on the other hand, agrees
with the prompt photon spectrum if its absolute value is normalized to the
data, corresponding to a large effective $K$-Factor of 5. 

Huovinen et al., fixing
the initial conditions ($T_0=210 - 250$ MeV) in their hydrodynamical model
partly by a comparison with hadron spectra,
were able to describe the data equally well with or without a phase
transition \cite{Huovinen00} (see Fig. \ref{fig4.8}).
They were able to fit the WA98 data without the need of 
an extremely high initial
temperature, an initial radial velocity, 
or in-medium hadron masses. This might be caused partly by a strong flow
at later stages since they do not assume a boost-invariant
longitudinal expansion. \footnote{There remains an apparent 
discrepancy in the results and conclusions
between the work by Srivastava and Sinha \cite{Srivastava00} and
by Huovinen and Ruuskanen \cite{Huovinen00} which is not yet 
sufficiently explained.}
\newpage

\clearpage

Summarizing, WA98 found a rather flat photon spectrum, 
which cannot be easily explained by conservative
models. It requires either a high initial
temperature, a large prompt photon contribution, an initial
radial velocity, in-medium modifications of the hadron masses
and/or a strong flow at later stages.
At the moment, it is fair to say that the uncertainties
and ambiguities in the
hydrodynamical models and in the rates do not allow to
decide from the WA98 photon spectra about the presence of a QCD phase
transition in SPS heavy-ion collisions. However, most calculations do require 
a thermal source with an initial temperature of $T_{i} \approx 250 \, 
\mathrm{MeV}$ or higher.

\section{Outlook for RHIC}

In summer 2000 experiments at the Relativistic Heavy Ion Collider 
(RHIC) at BNL started to take data in collisions of Au nuclei at 
$\sqrt{s_{NN}} = 130 \, \mathrm{GeV}$, continuing with a beam energy 
of $\sqrt{s_{NN}} = 200 \, \mathrm{GeV}$ from 2001 on. First 
results of the RHIC experiments have already been presented 
\cite{QM2001}, however results on direct photons are not available at 
this early stage.

One of the major goals of the PHENIX experiment \cite{daveQM97,billQM01} 
at RHIC is the measurement of direct photons in the central detector 
arms at midrapidity. Photon measurements and 
neutral meson reconstruction are performed 
with electromagnetic cal\-orimeters (EMCAL) using 
two different technologies, a lead glass detector, which consists of 
the transformed and updated calorimeter used in WA98 and a 
lead-scintillator sampling calorimeter. In addition, the sophisticated 
electron detection capabilities should also allow to measure inclusive 
photons via the e$^{+}$-e$^{-}$-pairs from conversions.
The central detectors cover $90^{\circ}$ in azimuth and the 
pseudorapidity range $ |\eta| < 0.35$. A central magnet provides an axial 
field, and tracking and momentum measurement is performed in 
three different sub-systems: pad chambers (PC), drift chambers (DC)
and time-expansion chambers (TEC). Electron identification is 
achieved by simultaneously using a ring imaging Cherenkov counter 
(RICH) for $p < 4.7 \, \mathrm{GeV}/c$, 
electromagnetic energy measurement in the calorimeters for
$p > 0.5 \, \mathrm{GeV}/c$ and 
$dE/dx$ measurement in the TEC for $p < 2 \, \mathrm{GeV}/c$. 
A planned upgrade of the TEC to a transition 
radiation detector (TRD) will further strengthen the electron 
identification. Photons converting in the outer shell of the 
multiplicity and vertex detector (MVD) can be identified as electron 
pairs with a small, but finite apparent mass\footnote{This finite 
mass is an artefact of the assumption of particle emission from the 
collision vertex.}. It is planned to add a converter plate to the 
experiment for part of the data taking to minimize uncertainties of 
the conversion probability and the location of the conversion point.
Photons with $p > 1 \, \mathrm{GeV}/c$ will be identified in the 
calorimeters with hadron suppression from the smaller deposited energy 
and additional rejection by time-of-flight (for slow hadrons) and shower 
shape analysis. Furthermore, charged hadrons will be identified by the 
tracking detectors in front of the calorimeters. The calorimeters will 
also measure $\pi^{0}$ and $\eta$ production necessary for the 
estimate of the decay photon background.

\begin{figure}[hbt]
\centering
\epsfbox{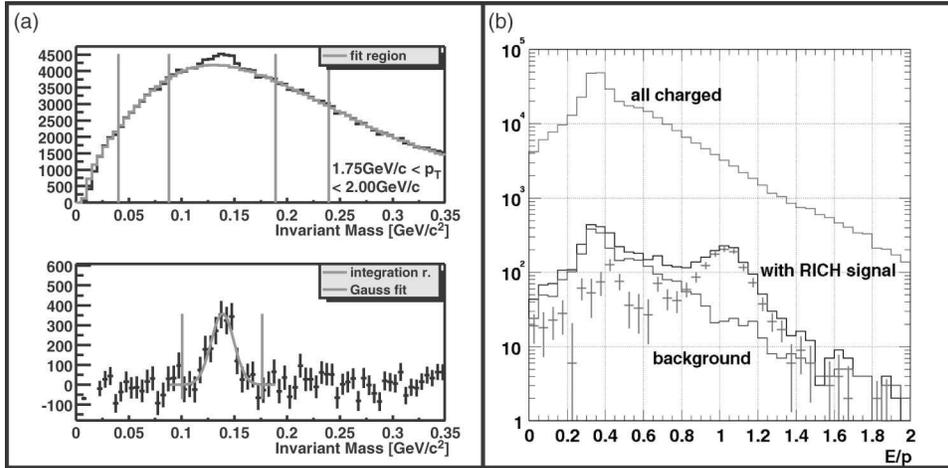}
\caption{a) Example of a two-photon invariant mass spectrum as 
measured by the PHENIX electromagnetic calorimeter (see e.g. 
\protect\cite{davidQM01}). b) Identification 
of electrons in the PHENIX experiment 
from the ratio of electromagnetic energy measured by 
calorimetry and momentum measured by tracking (see e.g. 
\protect\cite{akibaQM01}).
}
\protect\label{fig:phenix}
\end{figure}

Fig.~\ref{fig:phenix} illustrates the measurement capabilities of the 
PHENIX experiment already from the early data of the beam time in 2000. 
Fig.~\ref{fig:phenix}a) shows an invariant mass 
spectrum of photon pairs measured with the EMCAL. The peak of the 
neutral pions can easily be observed. In fact, transverse momentum 
spectra of neutral pions have already been extracted and published 
\cite{phenix:highpt}. With the higher statistics measurements in 2001 
at the full RHIC energy of $\sqrt{s_{NN}} = 200 \, \mathrm{GeV}$, also 
a measurement of the $\eta$ meson will be performed, providing the 
basis for an extraction of direct photons from the inclusive photon 
spectra. The high quality of the electron identification is shown in 
Fig.~\ref{fig:phenix}b), where distributions of the ratio of the 
calorimetric energy to the momentum are displayed. While for  
inclusive charged particles only a smooth distribution is observed, a 
clear peak at $E/p=1$ is observed when a signal of the RICH detector 
is required.

The different technologies should provide an excellent measurement of 
direct photons with independent checks of systematic errors. In 
addition, as RHIC is a dedicated heavy-ion accelerator, a much higher 
integrated luminosity is expected, which, together with the expected 
higher photon production rates, will make the RHIC measurements 
superior to the existing lower energy heavy-ion data.

The dynamic range of the photon measurements at RHIC should extend 
over the range $1.0 \, \mathrm{GeV}/c \le p_{T} \le 30 \, 
\mathrm{GeV}/c$, discrimination of high $p_{T}$ photons from merging 
$\pi^{0}$ should be possible up to $p_{T} = 25 \, \mathrm{GeV}/c$. 

\begin{figure}[hbt]
\centering
\epsfbox{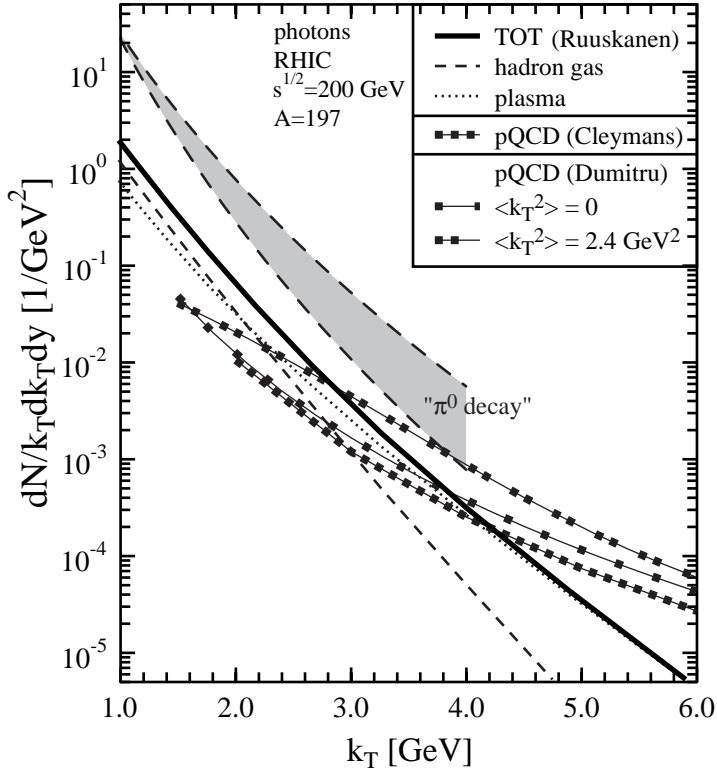}
\caption{Predictions for thermal and hard photon production at RHIC as 
discussed in the text compared to estimates of the background of 
inclusive photons from neutral pion decays.
}
\protect\label{fig:rhic}
\end{figure}

In Fig.~\ref{fig:rhic} predictions for direct photon production at 
RHIC are shown. Given are results of hydrodynamic calculations by 
Ruuskanen and R{\"a}s{\"a}nen \cite{Ruuskanen:lhc} assuming a QGP 
phase transition and an initial time of $\tau_0=0.17 
\,\mathrm{fm}/c$ and using the complete $\alpha_{S}$-resummed rates 
from Arnold et al. \cite{Arnold01}. 
The dashed line shows the contribution from the hadron 
gas and the dotted line the contribution from the QGP (resp. 
contributions during the mixed phase are included), while the solid 
line shows the sum of both contributions.
Also included are estimates of hard photon production from pQCD 
calculations by Cleymans et a. \cite{cleymans:hpc} and Dumitru et al. 
\cite{Dumitru01}. From this comparison one can see that the hard 
production should start to dominate the direct photon yield for 
transverse momenta larger than 3-4~GeV$/c$.

In addition, Fig.~\ref{fig:rhic} shows estimates of the background 
photons from neutral pion decays as a grey band. These are obtained by 
extrapolating results from the neutral pion measurements at 
$\sqrt{s_{NN}} = 130 \, \mathrm{GeV}$. The upper limit is calculated 
assuming a scaling of the pion yield according to equation 
\ref{eq:scaling}. The lower limit has been obtained by assuming that 
the pion spectrum at all transverse momenta scales as the total 
multiplicity density - for this scaling a value of 1.14 has been 
measured by the PHOBOS experiment \cite{phobos:mult200}. It can be 
seen that the direct photon production within the model used here may 
easily exceed a value of $10 \%$ of the pion decay photons and should 
thus be reliably measureable at RHIC. 

\section{Conclusions}
The first direct photon measurement in heavy ion reactions has been 
successfully performed by the WA98 experiment. The direct photon yield 
is higher than expected from simple extrapolations of earlier 
p-induced reactions. The results may be partially explained by an 
increased hard photon production in nuclei, i.e. $k_{T}$ broadening. 
The yield at intermediate $p_{T}$ seems to require a thermal source 
with a high initial temperature.

The future of direct photon measurements in heavy ion reactions looks 
bright in view of the advent of the RHIC experiments. The PHENIX 
experiment is especially geared up to measure photons, the redundancy 
in this experiment should provide a good control of systematic errors 
of these difficult measurements.

{\bf Acknowledgement:}
I would like to thank M.H. Thoma for valuable discussions and 
comments.


\begin{thebibliography}{99}
\bibitem{Alam96}
J. Alam, B. Sinha, and S. Raha, Phys. Rep. 273 (1996) 243.
\bibitem{Alam01}
J. Alam, S. Sarkar, P. Roy, T. Hatsuda, and B. Sinha, Ann. Phys. (N.Y.)
286 (2001) 159.
\bibitem{Pei02}
T. Peitzmann, and M.H. Thoma, accepted for publication in Phys. Rep. (2002), 
preprint hep-ph/0111114.
\bibitem{Aggarwal:2000}
Expt. WA98, M.M.~Aggarwal et al., Phys. Rev. Lett. 85 (2000) 3595.
\bibitem{Albrecht:1996}
Expt. WA80, R.~Albrecht et al., Phys. Rev. Lett. 76 (1996) 3506.
\bibitem{Albrecht:1991}
Expt. WA80, R.~Albrecht et al., Z. Phys. C 51 (1991) 1.
\bibitem{zpc:ake:90}
Expt. NA34 (HELIOS 2), T.~{\AA}kesson et al., Z. Phys. C 46 (1990) 369.
\bibitem{zpc:bau:96}
Expt. NA45 (CERES), R.~Baur, Z. Phys. C 71 (1996) 571.

\bibitem{Srivastava94}
D.K. Srivastava and B. Sinha, Phys. Rev. Lett. 73 (1994) 2421.
\bibitem{Shuryak94}
E.V. Shuryak and L. Xiong, Phys. Lett. B 333 (1994) 316.
\bibitem{Neumann95} 
J.J. Neumann, D. Seibert, and G. Fai, Phys. Rev. C 51 (1995) 1460.
\bibitem{Arbex95}
N. Arbex, U. Ornik, M. Pl\"umer, A. Timmermann, and R.M. Weiner,
Phys. Lett. B 345 (1995) 307.
\bibitem{Dumitru95}
A. Dumitru et al., Phys. Rev C 51 (1995) 2166.
\bibitem{Sollfrank97}
J. Sollfrank et al., Phys. Rev. C 55 (1997) 392.
\bibitem{Sarkar99}
S. Sarkar, P. Roy, J. Alam, and B. Sinha, Phys. Rev. C 60 (1999) 054907.
\bibitem{Cleymans97}
J. Cleymans, K. Redlich, and D.K. Srivastava, Phys. Rev. C 55 (1997) 1431.
\bibitem{Cleymans98}
J. Cleymans, K. Redlich, and D.K. Srivastava, Phys. Lett. B 420 (1998) 261.
\bibitem{Srivastava00a}
D.K. Srivastava and B. Sinha, Eur. Phys. J. C 12 (2000) 109 and
Erratum Eur. Phys. J. C 20 (2001) 397.

\bibitem{plb:ada:95}
Expt. E704, D.~Adams et al., Phys. Lett. 345B (1995) 569.
\bibitem{prl:mcl:83}
Expt E629, M.~McLaughlin et al., Phys. Rev. Lett. 51 (1983) 971.
\bibitem{zpc:bad:86}
Expt. NA3, J.~Badier et al., Z. Phys. C 31 (1986) 341.
\bibitem{rmp:owens:87}
J.F.~Owens, Rev. Mod. Phys. 59 (1987) 465.

\bibitem{Wong98}
C. Wong and H. Wang, Phys. Rev. C 58 (1998) 376.
\bibitem{Dumitru01}
A. Dumitru, L. Frankfurt, L. Gerland, H. St\"ocker, and M. Strickman,
Phys. Rev. C 64 (2001) 054909.

\bibitem{Srivastava00}
D.K. Srivastava and B. Sinha, Phys. Rev. C 64 (2001) 034902.
\bibitem{Alam01a}
J. Alam, S. Sarkar, T. Hatsuda, T.K. Nayak, and B. Sinha,
Phys. Rev. C 63 (2001) 021901.
\bibitem{Peressounko00a}
D.Y. Peressounko and Y.E. Pokrovsky, {\it hep-ph/0009025}.
\bibitem{Gallmeister00}
K. Gallmeister, B. K\"ampfer, and O.P. Pavlenko, Phys. Rev. C 62 (2000) 057901.
\bibitem{Huovinen00}
P. Huovinen, P.V. Ruuskanen, and S.S. R\"as\"anen, {\it nucl-th/0111052}.
\bibitem{Chaudhuri00}
A.K. Chaudhuri, {\it nucl-th/0012058}.
\bibitem{Steffen01}
F.D. Steffen and M.H. Thoma, Phys. Lett. B 510 (2001) 98.

\bibitem{Alam-icpaqgp}
J. Alam, these proceedings.

\bibitem{QM2001}
Quark Matter 2001, Proceedings of the Fifteenth International Conference
on Ultra-Relativistic Nucleus-Nucleus Collisions, Stony Brook, USA,
2001, to be published in Nucl. Phys. A.
\bibitem{daveQM97}
PHENIX Collaboration, D.~Morrison, {\it et al.}, Nucl. Phys. A {638}
(1998) 565c.
\bibitem{billQM01}
PHENIX Collaboration, W.~Zajc, {\it et al.}, in: Proceedings of the Quark
Matter Conference 2001, Nucl. Phys. A (2001) to be printed,
{\it nucl-ex/0106001}.

\bibitem{davidQM01}
PHENIX Collaboration, G.~David, {\it et al.}, in: Proceedings of the Quark
Matter Conference 2001, Nucl. Phys. A (2001) to be printed,
{\it nucl-ex/0105014}.
\bibitem{akibaQM01}
PHENIX Collaboration, Y.~Akiba, {\it et al.}, in: Proceedings of the Quark
Matter Conference 2001, Nucl. Phys. A (2001) to be printed.

\bibitem{phenix:highpt}
PHENIX Collaboration, K. Adcox et al., Phys. Rev. Lett. 88 (2002) 022301.

\bibitem{Ruuskanen:lhc}
P.V. Ruuskanen and S. R{\"a}s{\"a}nen, private communication and talk
presented at the Workshop: Hard Probes in HIC at the LHC, 
CERN, 2001.

\bibitem{Arnold01}
P. Arnold, G.D. Moore and L.G. Yaffe, 
{\it hep-ph/0109064} and {\it hep-ph/0111107}.

\bibitem{cleymans:hpc}
Hard Probe Collaboration, J. Cleymans et al., Int. J. Mod. Phys. 10 (1995) 2491.

\bibitem{phobos:mult200}
PHOBOS Collaboration, B. Back et al., submitted to Phys. Rev. 
Lett., {\it nucl-ex/0108009}.

\end{thebibliography}
\end{document}